\documentstyle[pre,multicol,aps]{revtex}

\begin{document}

\input{epsf.tex}
\epsfverbosetrue

\title{Stable vortex and dipole vector solitons in 
a saturable nonlinear medium}

\author{Jianke Yang$^\dagger$ and 
Dmitry E. Pelinovsky$^{\dagger\dagger}$}
\address{$^{\dagger}$
Department of Mathematics and Statistics, 
University of Vermont, Burlington, VT 05401, USA  \\
$^{\dagger\dagger}$ Department of Mathematics, 
McMaster University, Hamilton, Ontario, Canada, L8S 4K1} 
\maketitle

\begin{abstract}
We study both analytically and numerically the existence, 
uniqueness, and stability of vortex and dipole 
vector solitons in a saturable nonlinear medium in (2+1) dimensions.   
We construct perturbation series expansions 
for the vortex and dipole vector solitons near the bifurcation 
point where the vortex and dipole components are {\em small}. 
We show that both solutions uniquely bifurcate from 
the same bifurcation point. We also prove that 
both vortex and dipole vector solitons are linearly 
{\em stable} in the neighborhood of the bifurcation point. 
Far from the bifurcation point, the family of vortex solitons 
becomes linearly unstable via oscillatory instabilities, while 
the family of dipole solitons remains stable in the entire 
domain of existence. In addition, we show that an unstable vortex 
soliton breaks up either into a rotating dipole soliton or 
into two rotating fundamental solitons. 
\end{abstract}

\vspace{0.5cm}


\newpage

\section{Introduction}

Spatial solitons have been a subject of many studies 
since their first theoretical prediction \cite{chiao}. 
The previous research on spatial solitons was driven by 
their promising applications in all-optical devices 
in which the light guides and steers the light itself 
\cite{SpatialBook}. Early works studied 
optical materials with the Kerr (cubic) nonlinearity 
which exhibit stable fundamental (single-hump) solitons 
in one spatial dimension \cite{ZaSh:71,barthelemy,aitchison} 
and collapse in two and three spatial dimensions \cite{zakharov2,Berge}. 
Later works focused on optical materials with saturable nonlinear 
response such as photorefractive crystals (see \cite{segev} and references therein). 
The nonlinearity saturation suppresses the collapse of 
fundamental solitons in two and three 
dimensions \cite{soto2,enns,segev2}, which opens the door for 
their experimental observation in multi-dimensional optical 
beams. The instability of higher-order (multi-hump) solitons 
is not suppressed by the nonlinearity saturation however
\cite{grantham,edmundson,book,firthskryabin,Yang02}. Internal modes of 
fundamental solitons in saturable optical materials have also 
been reported \cite{Yang02,rosanov}. These modes are responsible for long-lived shape 
oscillations.

Recently, incoherent coupling of spatial solitons in photorefractive crystals
was proposed and experimentally demonstrated
\cite{segev2,anastassiou}. The mutual trapping 
of incoherent optical beams leads to many novel
spatial solitons such as vortex and dipole vector solitons 
\cite{buryak,ziad00,Muss,malmberg,garcia,carmon,kivshar00,ziad,musslimani,desyatnikov}. 
The incoherently coupled spatial solitons are described by a system of 
coupled nonlinear Schr\"{o}dinger (NLS) equations. 
A similar system of equations also describes temporal solitons 
in birefringent optical fibers 
and wavelength-division-multiplexed systems
\cite{menyuk,agrawal,hasegawa,ostrov00,YangPhysicaD,YangTanPRL}. 
Additionally, vortex vector solitons are known in the    
Bose-Einstein condensation guided by a magnetic trap 
\cite{skryabin}. 

Vortex and dipole vector solitons in saturable optical 
materials are interesting for both physical and mathematical 
reasons. Physically, these spatial solitons are novel nonlinear
objects. They bifurcate from 
a coupled state where a fundamental soliton in one component 
guides a small higher-order mode in the other component. 
Far from the bifurcation threshold, both components strongly 
trap each other and form a fully coupled vector soliton.   
Mathematically, existence and stability of the vortex and 
dipole vector solitons are challenging problems due to their 
complexity. The existence of vortex and dipole solitons was 
established numerically \cite{ziad00,garcia} and with a 
heuristic variational method \cite{desyatnikov}. However, analytical 
expressions for radial-angular dependences of vortex and 
dipole solitons have not been found. 
On the stability of vortex solitons, the numerical results of \cite{ziad00}
suggest that vortex solitons with {\em small} vortex components are stable 
and observable for large propagation distances, 
while vortex solitons with large vortex components are unstable. 
Numerical results of \cite{garcia} show however that 
{\em all} vortex solitons are linearly unstable, and the 
instability leads to the breakup of vortex solitons 
into rotating dipole solitons. The discrepancy between 
\cite{ziad00} and \cite{garcia} raises an open question: 
are vortex vector solitons with small vortex components really stable or not? 
On the other hand, dipole solitons were found numerically to be always stable in \cite{garcia}. 

In this paper, we clarify the issues of existence and 
stability of vortex and dipole vector solitons 
in a saturable nonlinear medium such as photorefractive crystals. 
First, we address the existence and uniqueness of vortex and dipole 
solitons with the perturbation technique \cite{dmitryYang,PelKiv}.
We derive perturbation series expansions for 
the vortex and dipole solitons near the bifurcation point where
the vortex and dipole components are {\em small}. 
The analytical formulas for
these solitons are in excellent agreement with our numerical results. 
We also prove that those vector solitons are 
unique up to phase-, translation-, and rotation-invariances.
Next, we study the linear stability of vortex and 
dipole solitons with both the spectral analysis and numerical methods.  
We show that dipole solitons are linearly stable 
in the entire existence domain, while the 
vortex solitons with {\em large} vortex components 
are linearly unstable, in agreement with \cite{garcia}. 
However, we prove that vortex solitons with 
{\em small} vortex components are linearly 
{\em stable}, confirming the results of \cite{ziad00}, not \cite{garcia}. 
Lastly, we study the nonlinear evolution of linearly 
unstable vortex solitons. 
We show that an unstable vortex soliton 
breaks up into a rotating dipole soliton only when the vortex component is 
below a certain threshold. Above this threshold, an unstable vortex soliton
breaks up into two fundamental vector solitons instead. 

\section{Existence and uniqueness of vortex and dipole solitons}

The mathematical model for two incoherently-coupled 
laser beams in a photorefractive crystal is well known 
(see, e.g., \cite{ziad00,garcia}). After variable
rescalings, the model can be written as 
a system of coupled equations:
\begin{equation} \label{E1}
i\frac{\partial E_1}{\partial z} + \Delta E_1 +
\frac{E_1(|E_1|^2+|E_2|^2)}{1+s(|E_1|^2+|E_2|^2)}=0,
\end{equation}
\begin{equation} \label{E2}
i\frac{\partial E_2}{\partial z}+\Delta E_2 +
\frac{E_2(|E_1|^2+|E_2|^2)}{1+s(|E_1|^2+|E_2|^2)}=0,
\end{equation}
where $\Delta$ is the two-dimensional 
Laplacian, and $s$ is the saturation parameter. 

Vector solitons in this model take the form:
\begin{equation}
E_1=u(x,y)e^{iz}, \quad E_2=w(x,y)e^{i\lambda z},
\end{equation}
where the frequency of the $E_1$ wave has been normalized 
to one, and the frequency of the $E_2$ wave is $\lambda$. 
The amplitude functions $u(x,y)$ and $w(x,y)$ satisfy 
the nonlinear boundary-value problem with zero 
boundary conditions on the $(x,y)$ plane:
\begin{equation} \label{u}
\Delta u-u+\frac{u(|u|^2+|w|^2)}{1+s(|u|^2+|w|^2)}=0, 
\end{equation}
\begin{equation} \label{w}
\Delta w-\lambda w+\frac{w(|u|^2+|w|^2)}{1+s(|u|^2+|w|^2)}=0. 
\end{equation}

The system (\ref{u})--(\ref{w}) may have several types of 
vector solitons localized in two dimensions. 
The {\em fundamental} vector solitons take the form:
\begin{equation}
\label{fundamental}
u = c_u \Phi(r), \;\;\;\;
w = c_w \Phi(r), 
\end{equation}
where $\Phi(r)$ is a real-valued, single-hump function, 
$r = \sqrt{x^2 + y^2}$, and $c_u$ and $c_w$ are 
arbitrary complex parameters
constrained by the relation $|c_u|^2+|c_w|^2=1$. 
These solitons exists at $\lambda = 1$ 
where the system (\ref{u})--(\ref{w}) reduces to a scalar equation 
for $\Phi(r)$, see Eq. (\ref{u0}) below. The {\em vortex} vector 
solitons take a general form:
\begin{equation}
\label{vortex}
u= \Phi_u(r) \; e^{i n \theta}, \quad 
w= \Phi_w(r) \; e^{i m \theta},
\end{equation}
where $n$ and $m$ are topological charges of vortices 
in the $u$ and $w$ components, $\Phi_u(r)$ and $\Phi_w(r)$ 
are real-valued functions, and $(r, \theta)$ are the 
polar coordinates on the $(x, y)$ plane. The simplest 
vortex soliton has $n=0$ and $m=1$
\cite{ziad00,garcia}.
The {\em multi-pole} vector solitons
take yet another general form:
\begin{equation}
\label{dipole}
u = U(r,\theta), \;\;\;\;
w = W(r,\theta),
\end{equation}
where $U(r,\theta)$ and $W(r,\theta)$ are real-valued functions
and may have multi-hump profiles on the $(x, y)$ plane. 
The multi-pole solitons with a single hump for $u(x,y)$ 
and multiple humps for $w(x,y)$ were approximated 
in the variational approach \cite{desyatnikov} 
by the ansatz:
\begin{equation} \label{ansatz}
u = U(r), \quad 
w = W(r) \cos m \theta,
\end{equation}
where the number of humps in the $w$ component is $2 m$. 
The simplest multi-pole soliton is a dipole soliton which has $m = 1$.

In this paper, we study the simplest vortex and multi-pole vector solitons. 
We will refer to them simply as the vortex soliton and 
dipole soliton hereafter. 
The existence and uniqueness of the vortex and dipole solitons 
can be studied by perturbation methods. 
The perturbation series expansions are derived 
in the neighborhood of the bifurcation value 
$\lambda = \lambda_0(s)$, where the $w$ component 
is small. With the perturbation arguments, we 
show that the bifurcation value $\lambda_0(s)$ 
is the same for both branches of 
vortex and dipole solitons, and these solitons 
bifurcate uniquely from $\lambda=\lambda_0(s)$.  

\begin{figure}
\setlength{\epsfxsize}{8cm}\epsfbox{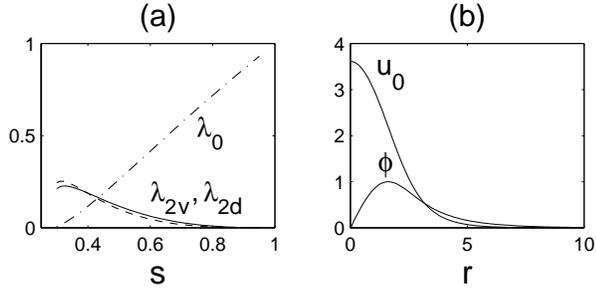}
\caption{(a) The cut-off frequency $\lambda_0$ (dashed-dotted) and 
the correction terms $\lambda_{2v}$ (dashed) and $\lambda_{2d}$ (solid) 
for vortex and dipole solitons as a function of $s$. (b) The scalar $u_0(r)$ soliton 
and the normalized eigenfunction $\phi(r)$ at $s=0.5$.}
\label{lambda02}
\end{figure}

Setting $w=0$ in (\ref{u}), we find the nonlinear 
boundary-value problem for the scalar soliton $u = u_0(r)$: 
\begin{equation} \label{u0}
u_{0}'' + \frac{1}{r}u'_{0} - u_0+\frac{u_0^3}{1+su_0^2}=0, 
\end{equation}
where $u_0(r)$ is a real-valued function. 
We take $u_0(r)$ to be the fundamental soliton, i.e., $u_0(r)>0$ 
for finite $r \geq 0$.  
If $w$ is small in (\ref{w}), we get a linear 
eigenvalue problem for the first-order correction 
$w = w_1(x,y)$:
\begin{equation} \label{eigen}
\Delta w_1 -\lambda w_1 + \frac{u_0^2}{1+su_0^2} w_1 = 0, 
\end{equation}
where $w_1(x,y)$ is a complex-valued function in general and $\lambda$ is 
the eigenvalue. The linear equation
(\ref{eigen}) supports localized solutions of 
the form $\phi(r) e^{\pm i m \theta}$ 
at a set of discrete values of $\lambda$. Since 
we study the simplest 
vortex and dipole solitons, we set $m=1$, and 
require $\phi(r)$ to be a non-negative function for $r \geq 0$.  
The corresponding eigenvalue $\lambda_0(s)$ and eigenfunction 
$\phi(r)$ satisfy the following reduced equation:
\begin{equation} \label{phi}
\phi'' + \frac{1}{r}\phi' - \left( \lambda_0+
\frac{1}{r^2} \right) \phi + 
\frac{u_0^2}{1+su_0^2}\phi=0.
\end{equation}
The eigenvalue $\lambda$ is unique once $s$ is fixed. 
We normalize the eigenfunction $\phi(r)$ such that it 
has a maximum value one. Numerically, we compute $\lambda_0(s)$ and $\phi(r)$ by 
the shooting method. Fig. \ref{lambda02}a shows 
the dependence of $\lambda_0$ versus saturation 
parameter $s$. Fig. \ref{lambda02}b shows the scalar 
soliton $u_0(r)$ and the normalized eigenfunction 
$\phi(r)$ at $s=0.5$, where $\lambda_0 = 0.2622$. 

When $\phi(r)$  and $\lambda_0(s)$ are known, a general solution for $w_1(r,\theta)$ 
can be written in the form:
\begin{equation} \label{w1formula}
w_1(r,\theta)=\phi(r)(\cos\theta+ip\sin\theta), 
\end{equation}
where $p$ is an arbitrary real parameter. 
In this general solution, we have removed 
arbitrary rotations and translations on the $(x,y)$ plane
as well as an arbitrary phase shift in the $w$ 
component. We note that the
solution (\ref{w1formula}) is identical to 
the variational ansatz in \cite{desyatnikov}.

Below, we use the perturbative method and show 
that there are only two continuations of 
the solution (\ref{w1formula}): for 
$p = \pm 1$ and $p = 0$. When $p=\pm 1$, 
the perturbation series expansion recovers 
the vortex soliton (\ref{vortex}) with 
$n = 0$ and $m = \pm 1$. When $p=0$, the perturbation 
expansion recovers the dipole soliton (\ref{dipole}) 
with a single hump for $u$ and a double hump for $w$. 
For given values of $s$ and $\lambda$, the two solutions 
are unique up to phase-, translation- and 
rotation-invariances. At other values of $p$, the solution with $w$'s 
leading-order term as (\ref{w1formula}) can not exist. 

The perturbation series expansions for 
vector solitons in system (\ref{u}) and (\ref{w}) 
take the form:
\begin{equation} \label{uexpand}
u=u_0(r) + \epsilon^2 u_2(r,\theta) + 
\epsilon^4 u_4(r,\theta) + {\rm O}(\epsilon^6), 
\end{equation}
\begin{equation}  \label{wexpand}
w=\epsilon w_1(r,\theta) + \epsilon^3 w_3(r,\theta) 
+ \epsilon^5 w_5(r,\theta) + {\rm O}(\epsilon^7), 
\end{equation}
and 
\begin{equation} \label{lambdaexpand}
\lambda=\lambda_0(s) + \epsilon^2 \lambda_2(s) + 
\epsilon^4 \lambda_4(s) + {\rm O}(\epsilon^6),
\end{equation}
where $\epsilon$ is a small parameter, 
$u_0(r)$ is the scalar fundamental soliton 
solving (\ref{u0}), $w_1(r,\theta)$ is the first-order 
correction in the form (\ref{w1formula}), and the 
cut-off frequency $\lambda_0$ is the eigenvalue 
of (\ref{phi}). The objective of 
the perturbation analysis is to uniquely determine 
the coefficients $\lambda_2, \lambda_4, \dots$ 
as well as expressions for functions $u_2, u_4, w_3, w_5$ 
and so on. Once the coefficients $\lambda_0, \lambda_2, \dots$ 
have been obtained, we can compute $\epsilon$ from 
$\lambda$ in the expansion (\ref{lambdaexpand}). 
Once $\epsilon$ is found, together with functions 
$u_0, u_2, w_1, w_3, \dots$, we can approximate 
the vector soliton by the expansions (\ref{uexpand}) 
and (\ref{wexpand}). Below, we will carry out the 
perturbative calculations to order $\epsilon^3$. 

Substituting the perturbation series (\ref{uexpand}), 
(\ref{wexpand}) and (\ref{lambdaexpand}) into the 
original equations (\ref{u}) and (\ref{w}), 
at order $\epsilon^2$, we get an inhomogeneous equation for $u_2$: 
\begin{equation} \label{u2}
\Delta u_2 - u_2 + \frac{u_0^2(2 + s u_0^2)}{(1+su_0^2)^2} u_2 
+ \frac{u_0^2}{(1+su_0^2)^2} \bar{u}_2 = 
-\frac{u_0 |w_1|^2}{(1+su_0^2)^2}.
\end{equation}
where $\bar{u}$ is the complex conjugate of $u$.  
The linearized operator in the left-hand-side of 
(\ref{u2}) has a non-empty null space spanned by 
three linearly independent localized eigenfunctions:
\begin{equation}
\label{zeroeigenfunction}
u_{2h}^{(1)}=iu_0(r), \quad 
u_{2h}^{(2)}=u'_{0}(r) \cos\theta, \quad 
u_{2h}^{(3)}=u'_{0}(r) \sin\theta. 
\end{equation}
These eigenfunctions correspond to the phase 
and translational invariances of solitons in 
the scalar $u$ equation. The right-hand-side 
of (\ref{u2}) is orthogonal to $u_{2h}^{(1)}$ 
because it is real-valued. It is also orthogonal 
to $u_{2h}^{(2)}$ and $u_{2h}^{(3)}$ because 
it has different angular dependence of $1$ and 
$\cos 2 \theta$ (rather than $\cos \theta$ and $\sin \theta$). 
Therefore, up to phase-, translation-, and rotation-shifts, 
a localized solution to (\ref{u2}) is constructed uniquely 
in the form:
\begin{equation}\label{u2formula}
u_2= \frac{1}{2} (1+p^2)u_{20}(r) + 
\frac{1}{2}(1-p^2) u_{22}(r)\cos 2\theta, 
\end{equation}
where functions $u_{20}(r)$ and $u_{22}(r)$ satisfy the equations
\begin{equation} \label{u20}
u_{20}''+\frac{1}{r}u_{20}'-u_{20}+
\frac{u_0^2(3 + s u_0^2)}{(1+su_0^2)^2}u_{20}
=-\frac{u_0\phi^2}{(1+su_0^2)^2}, 
\end{equation}
\begin{equation} \label{u22}
u_{22}''+\frac{1}{r}u_{22}'-\left(1+\frac{4}{r^2}\right) u_{22}+
\frac{u_0^2(3 + s u_0^2)}{(1+su_0^2)^2}u_{22}  = 
-\frac{u_0\phi^2}{(1+su_0^2)^2}.
\end{equation}
We do not know exact analytical expressions for $u_{20}(r)$ 
and $u_{22}(r)$ but can compute them numerically.  

At order $\epsilon^3$, we get the equation for $w_3$ as 
\begin{equation} \label{w3old}
\Delta w_3 -\lambda_0 w_3 + \frac{u_0^2}{1+su_0^2} w_3 
=\left\{\lambda_2 - \frac{|w_1|^2+2u_0u_2}{(1+su_0^2)^2}\right\}w_1.
\end{equation}
We denote 
\begin{equation} \label{h12}
h_1(r)=\frac{\phi^2+2u_0u_{20}}{2 (1+su_0^2)^2}, \quad 
h_2(r)=\frac{\phi^2+2u_0u_{22}}{2 (1+su_0^2)^2}, 
\end{equation}
and rewrite Eq. (\ref{w3old}) in the form:
\begin{eqnarray} \label{w3}
\Delta w_3 -\lambda_0 w_3 + \frac{u_0^2}{1+su_0^2} w_3 
= \left[ \lambda_2-(1+p^2)h_1- \frac{1}{2} (1-p^2)h_2\right] 
\phi \cos\theta  \nonumber \\ 
+ i p \left[\lambda_2-(1+p^2)h_1+ \frac{1}{2} (1-p^2)h_2\right] 
\phi \sin\theta  - \frac{1}{2} (1-p^2)h_2\phi\cos 3\theta - 
\frac{1}{2} i p(1-p^2)h_2\phi\sin 3\theta.
\end{eqnarray}
The homogeneous part in (\ref{w3}) supports 
two linearly independent localized solutions 
$\phi(r)\cos\theta$ and $\phi(r)\sin\theta$. 
As a result, a localized solution of the non-homogeneous 
equation (\ref{w3}) exists if and only if the following 
solvability conditions are satisfied: 
\begin{equation} \label{solva1}
\int_{0}^\infty r\phi^2 \left\{\lambda_2-(1+p^2)h_1(r) 
- \frac{1}{2} (1-p^2)h_2(r)\right\}dr=0, 
\end{equation}
\begin{equation} \label{solva2}
p \int_{0}^\infty r\phi^2 \left\{\lambda_2-(1+p^2)h_1(r) 
+ \frac{1}{2} (1-p^2)h_2(r) \right\}dr=0.
\end{equation}
We will show below that 
these solvability conditions define only two 
perturbation series solutions for vector solitons. 
These solutions correspond to the choice $p = \pm 1$ 
or $p = 0$, which produce vortex and dipole 
solitons respectively.  

\subsection{Vortex solitons}

If $p\ne 0$, we eliminate parameter $\lambda_2$ from 
the system (\ref{solva1})--(\ref{solva2}) and find 
the solvability condition in the form:
\begin{equation} \label{solva3}
(1-p^2)\int_0^\infty r\phi^2 h_2(r) dr=0. 
\end{equation}
The integral in Eq. (\ref{solva3}) only depends 
on the parameter $s$, not on $p$. 
We have checked numerically that this integral 
never vanishes for any $s$. Thus, the condition (\ref{solva3}) 
is satisfied only when $p=\pm 1$. In this case, 
it follows from (\ref{w1formula}), (\ref{u2formula}) 
and (\ref{w3}) that $w_1=\phi(r) e^{\pm i \theta}$, 
$u_2 = u_{20}(r)$, and $w_3 = f(r)e^{\pm i \theta}$. 
We can continue the perturbation series expansions 
(\ref{uexpand})--(\ref{lambdaexpand})
to higher orders and find that all $u_{2n} \; (n\ge 0)$ corrections
are only functions of $r$, and all $w_{2n+1} \; (n\ge 0)$ corrections
have the form $g(r)e^{\pm i \theta}$. Thus, the perturbation series 
solution gives a vortex vector soliton (\ref{vortex}) with $n=0$ and $m=\pm 1$.  

\begin{figure}
\begin{tabular}{cc}
\setlength{\epsfxsize}{5cm}\epsfbox{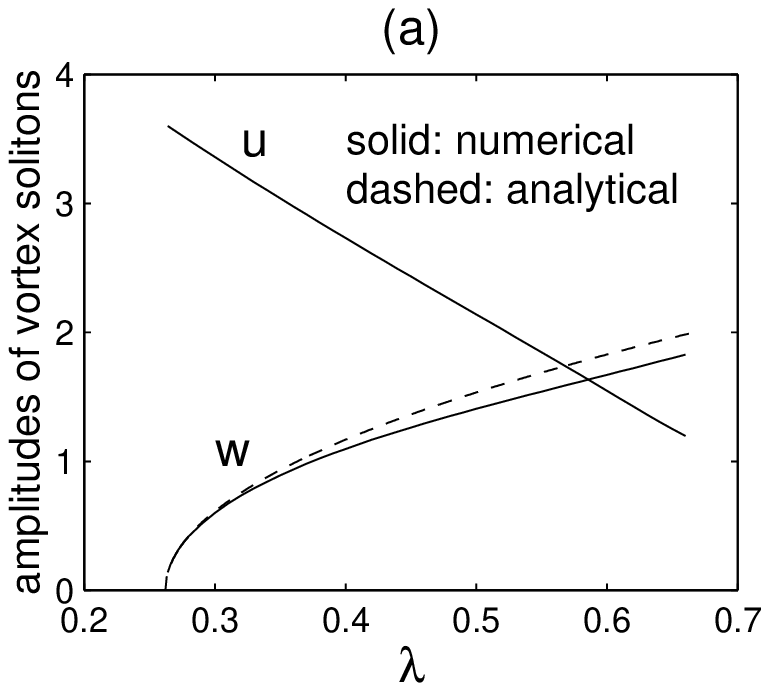}
\setlength{\epsfxsize}{3cm}\epsfbox{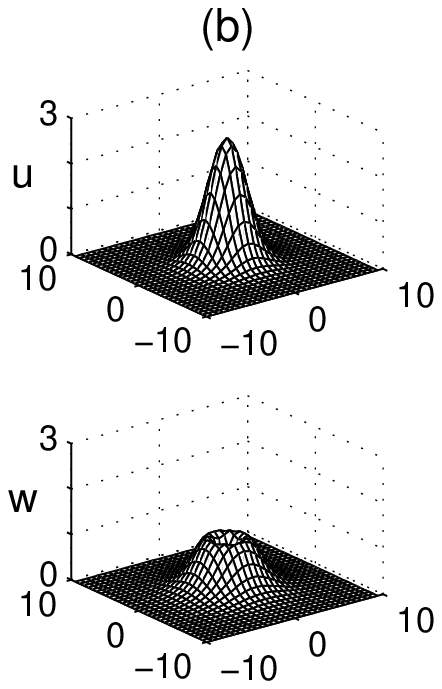}
\end{tabular}
\caption{(a) Amplitudes of the vortex vector solitons obtained analytically
(dashed line) and numerically (solid line) for $s=0.5$ and various frequencies 
$\lambda$. (b) A numerical vortex-soliton solution with $s=0.5$ and $\lambda=0.4$.}
\label{vortexfig}
\end{figure}

When $p=\pm 1$, we find from Eq. (\ref{solva1}) that the coefficient $\lambda_2$ is 
\begin{equation} \label{vortexlambda2}
\lambda_2=\lambda_{2v}(s) \equiv \frac{2 \int_0^\infty r \phi^2 h_1 dr}{\int_0^\infty r \phi^2 dr}.
\end{equation}
The functional dependence of $\lambda_{2v}$ versus $s$ 
is computed from this formula and plotted in Fig. \ref{lambda02}a (dashed line)
alongside the cut-off frequency $\lambda_0(s)$. 
Since the (non-negative) function $\phi(r)$ is 
normalized to have maximum one, the perturbation 
parameter $\epsilon$ determines the amplitude 
(maximum) of the vortex component $w$ with error at the  
order of $\epsilon^3$, see (\ref{w1formula}) and 
(\ref{wexpand}). The dependence of $\epsilon$ versus 
$\lambda$ and $s$ can be obtained from the perturbation 
series (\ref{lambdaexpand}) with error of order $\epsilon^2$:
\begin{equation} \label{vortexepsilon}
\epsilon = \sqrt{\frac{\lambda-\lambda_0(s)}{\lambda_{2v}(s)}}. 
\end{equation}
We compare the analytical formula (\ref{vortexepsilon}) 
with numerical results for $s=0.5$, where $\lambda_0=0.2622$ 
and $\lambda_{2v}=0.1010$. A dashed line in Fig. \ref{vortexfig}(a) shows 
the amplitude of the vortex component $w$ computed from 
(\ref{vortexepsilon}). 
Numerically, vortex solitons are computed from the original system 
(\ref{u})--(\ref{w}) by the shooting method. 
The amplitudes of the $u$ and $w$ components are also shown in 
Fig. \ref{vortexfig}. In Fig. \ref{vortexfig}(b),  
a profile of $u(x,y)$ and $w(x,y)$ components 
for $s = 0.5$ and $\lambda = 0.4$ is shown. 
We can see from Fig. \ref{vortexfig}a that 
the agreement between the analytical predictions and numerical values on the 
amplitudes of the $w$-component 
is good over a wide range of $\lambda$ values. 
Also, the numerically-obtained amplitude of the $u$-component depends 
linearly on $\lambda$, which is in agreement with 
the perturbation series (\ref{uexpand}) up to error of order $\epsilon^4$, 
since $\epsilon^2 \propto (\lambda-\lambda_0)$. 

\subsection{Dipole solitons}

If $p=0$, the condition (\ref{solva2}) is satisfied, 
while the condition (\ref{solva1}) gives the correction 
term $\lambda_2$ in the form:
\begin{equation} \label{lambda2}
\lambda_2=\lambda_{2d}(s) \equiv \frac{\int_0^\infty r \phi^2 
( 2 h_1 + h_2 ) dr}{2 \int_0^\infty r \phi^2 dr}.
\end{equation}
The functional dependence of $\lambda_{2d}$ versus $s$ 
is numerically computed and plotted in Fig. \ref{lambda02}(a) (solid line). 
When $\lambda_2$ is given 
by (\ref{lambda2}), a localized solution $w_3(r,\theta)$ of Eq. (\ref{w3}) 
exists, and this solution is real-valued. 
We can further show that the perturbation series 
expansions (\ref{uexpand})--(\ref{lambdaexpand})
can be successfully continued to higher orders 
of $\epsilon$, and a dipole-soliton solution  
(\ref{dipole}) can be obtained. This solution is real-valued and has the 
symmetries 
\begin{eqnarray} \label{symmetry}
u(-x, y)= u(x, y), \quad u(x, -y)=u(x,y), \quad 
w(-x, y)=-w(x, y), \quad w(x, -y)=w(x,y).
\end{eqnarray} 

\begin{figure}
\begin{tabular}{cc}
\setlength{\epsfxsize}{5cm}\epsfbox{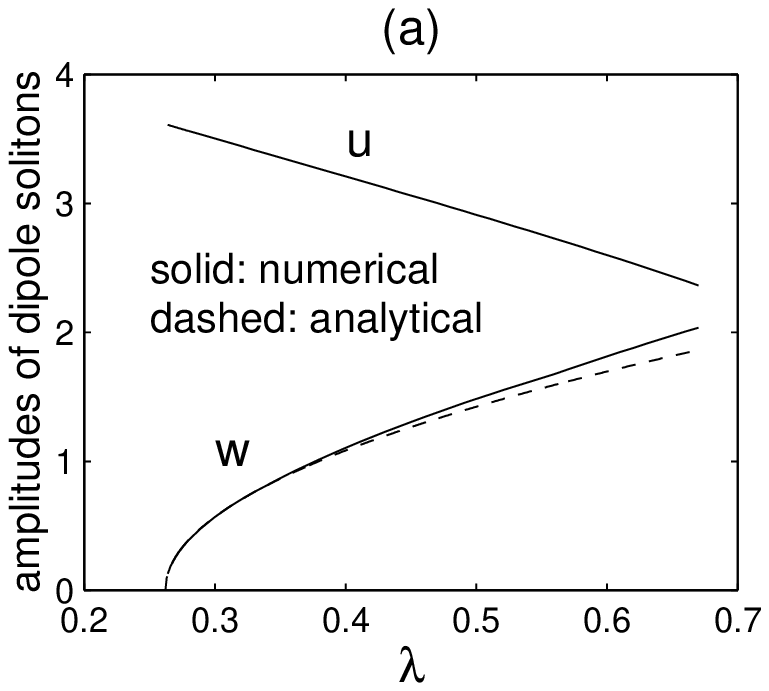}
\setlength{\epsfxsize}{3cm}\epsfbox{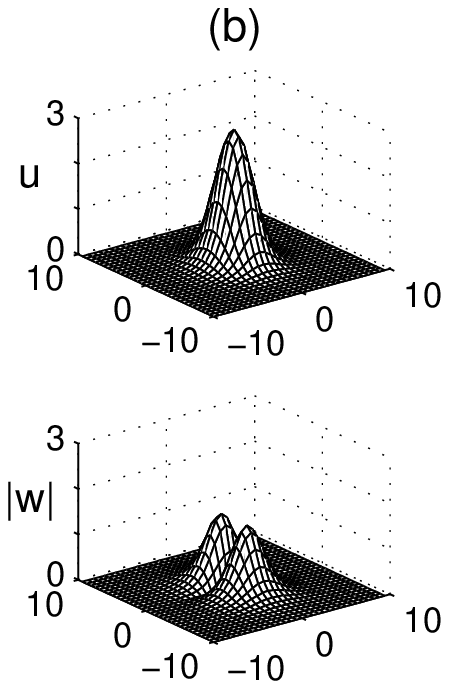}
\end{tabular}
\caption{(a) Amplitudes of the dipole vector solitons obtained analytically
(dashed line) and numerically (solid line) for $s=0.5$ and various frequencies $\lambda$.
(b) A numerical dipole-soliton solution with $s=0.5$ and $\lambda=0.5$.}
\label{dipolefig}
\end{figure}

Similar to the vortex soliton case, the perturbation 
parameter $\epsilon$ here gives the amplitude 
(maximum) of the dipole component $w$ with accuracy 
of O($\epsilon^3$). The formula for $\epsilon$ is still (\ref{vortexepsilon}), 
but the $\lambda_2$ value is now given by Eq. (\ref{lambda2}) instead of 
(\ref{vortexlambda2}). 
The comparison between the analytical result 
(\ref{vortexepsilon})--(\ref{lambda2})
and numerical results for dipole solitons is shown 
in Fig. \ref{dipolefig}a for $s=0.5$. In this case, 
$\lambda_0=0.2622$ and $\lambda_{2d}=0.1174$. 
We see again that the agreement between numerical 
and analytical results is very good
over a wide range of $\lambda$ values. In Fig. \ref{dipolefig}b, 
the profiles of $u(x,y)$ and $w(x,y)$ components of a dipole soliton, 
computed with numerical iteration methods, are displayed. 

We note that in view of Eqs. (\ref{vortexlambda2}) and (\ref{lambda2}), 
the integral in Eq. (\ref{solva3}) is actually 
\begin{equation} \label{integral}
\int_0^\infty r\phi^2 h_2(r) dr=(2\lambda_{2d}-\lambda_{2v})\int_0^\infty r\phi^2 dr. 
\end{equation}
Inspection of the $\lambda_{2d}(s)$ and $\lambda_{2v}(s)$ curves in Fig. \ref{lambda02}a
immediately confirms that the integral (\ref{integral}) is always positive. 
Thus Eq. (\ref{solva3}) holds only when $p=\pm 1$.

\vspace{0.25cm}

To summarize, we have shown that there are 
only two vector solitons of the system (\ref{u})--(\ref{w}) that
bifurcate from the cut-off frequency $\lambda = \lambda_0(s)$. 
They are either a vortex soliton (\ref{vortex}) 
or a real-valued dipole soliton (\ref{dipole}). 
The solutions are determined in terms of perturbation 
series expansions up to order $\epsilon^3$. 
Both solitons are unique up to phase, position 
and rotation invariances. The analytical 
results are confirmed by numerical calculations. 
Computations of the perturbation series expansions 
prove the existence and uniqueness of vortex and 
dipole vector solitons observed numerically in 
\cite{ziad00,garcia,desyatnikov}. 

\section{Linear stability of vortex and dipole solitons}

In this section, we study the linear stability of vortex and 
dipole vector solitons by spectral analysis, supplemented 
by numerical computations. Since the linearization operators 
differ for vortex and dipole solitons, we shall 
treat the two cases separately.

\subsection{Vortex solitons}
\label{section3a}

To study the linear stability of the 
vortex solitons (\ref{vortex}) with $n = 0$ and $m = 1$, 
we linearize the system (\ref{E1})--(\ref{E2}) with the 
perturbation in the form:
\begin{equation} \label{E1perturb}
E_1 = e^{iz} \left[ \Phi_u(r) + 
u_+(r) e^{- i n \theta + \sigma z} + 
\bar{u}_-(r) e^{i n \theta + \bar{\sigma} z} \right], 
\end{equation}
\begin{equation} \label{E2perturb}
E_2=e^{i\lambda z + i\theta} \left[ \Phi_w(r)+ 
w_+(r) e^{-in \theta + \sigma z} + 
\bar{w}_-(r) e^{i n \theta + \bar{\sigma} z} \right]. 
\end{equation}
The linearization problem can be written in the form:
\begin{eqnarray}
\label{linear1}
i \sigma u_+ & = & - u_+'' - \frac{1}{r} u_+' 
+ \left(1 + \frac{n^2}{r^2}\right) u_+ 
- V \; u_+ - V_{uu} \; (u_+ + u_-) - V_{uw} \; (w_+ + w_-), \\
\label{linear2}
-i \sigma u_- & = & - u_-'' - \frac{1}{r} u_-' 
+ \left(1 + \frac{n^2}{r^2}\right) u_- 
- V \; u_- - V_{uu} \; (u_+ + u_-) - V_{uw} \; (w_+ + w_-), \\
\label{linear3}
i \sigma w_+ & = & - w_+'' - \frac{1}{r} w_+' 
+ \left(\lambda + \frac{(n-1)^2}{r^2}\right) w_+ 
- V \; w_+ - V_{uw} \; (u_+ + u_-) - V_{ww} \; (w_+ + w_-), \\
\label{linear4}
-i \sigma w_- & = & - w_-'' - \frac{1}{r} w_-' 
+ \left(\lambda + \frac{(n+1)^2}{r^2}\right) w_- 
- V \; w_- - V_{uw} \; (u_+ + u_-) - V_{ww} \; (w_+ + w_-), 
\end{eqnarray}
where 
$$
V = \frac{\Phi_u^2 + \Phi_w^2}{1 + s(\Phi_u^2 + \Phi_w^2)}, \quad
V_{uu} = \frac{\Phi_u^2}{(1 + s(\Phi_u^2 + \Phi_w^2))^2}, \quad
V_{uw} = \frac{\Phi_u \Phi_w}{(1 + s(\Phi_u^2 + \Phi_w^2))^2}, \quad
V_{ww} = \frac{\Phi_w^2}{(1 + s(\Phi_u^2 + \Phi_w^2))^2}.
$$
The linearized problem can be formulated in the Hamiltonian form 
\cite{skryabin}:
\begin{equation}
\label{Lphi}
\pm i \sigma u_{\pm} = \frac{\delta h}{\delta \bar{u}_{\pm}}, \quad
\pm i \sigma w_{\pm} = \frac{\delta h}{\delta \bar{w}_{\pm}}, 
\end{equation}
where $h$ is the energy quadratic form associated 
with an eigenvector ${\bf u} = (u_+,u_-,w_+,w_-)^T$ 
and a linearized self-adjoint operator ${\cal L}$ 
of the right-hand-sides of the system (\ref{linear1})--(\ref{linear4}):
\begin{equation}
\label{quadform}
h = \langle {\bf u}, {\cal L} {\bf u} \rangle = 
i \sigma \int_0^{\infty} r dr \left( |u_+|^2 - |u_-|^2 + |w_+|^2 - 
|w_-|^2 \right). 
\end{equation} 
The eigenvalue $\sigma$ is defined by the spectrum of 
the linearized problem (\ref{linear1})--(\ref{linear4}) 
when ${\bf u}(r)$ is localized as $r \to \infty$ such 
that the integral (\ref{quadform}) makes sense. 
The eigenvalues could be isolated or embedded 
into continuous spectrum of the system (\ref{linear1})--(\ref{linear4}). 
The vortex soliton is linearly unstable if there exists 
an eigenvalue $\sigma$ for some $n$ such that ${\rm Re}(\sigma)>0$. 
We note that if $(\sigma, n, u_+, u_-, w_+, w_-)$ is a solution of  
the linear system (\ref{linear1})--(\ref{linear4}), so are 
$(\bar{\sigma}, -n, \bar{u}_-, \bar{u}_+, \bar{w}_-, \bar{w}_+)$,
$(-\sigma, -n, u_-, u_+, w_-, w_+)$ and $(-\bar{\sigma}, n, \bar{u}_+, \bar{u}_-, 
\bar{w}_+, \bar{w}_-)$. Thus, complex unstable eigenvalues $\sigma$
always come in quartets, while real and imaginary eigenvalues 
$\sigma$ always come in pairs. We also note that
eigenmodes $(\sigma, n, u_+, u_-, w_+, w_-)$ and 
$(\bar{\sigma}, -n, \bar{u}_-, \bar{u}_+, \bar{w}_-, \bar{w}_+)$
give the identical perturbation in (\ref{E1perturb}) and (\ref{E2perturb}), so 
do  $(-\sigma, -n, u_-, u_+, w_-, w_+)$  and
$(-\bar{\sigma}, n, \bar{u}_+, \bar{u}_-, \bar{w}_+, \bar{w}_-)$. 

According to the stability theory of solitary waves in 
the system of coupled NLS equations 
\cite{dmitry02}, only eigenvalues $\sigma$ with negative 
or zero values of the energy quadratic form (\ref{quadform})
may bifurcate to the domain ${\rm Re}(\lambda) > 0$, leading to instabilities. 
We shall apply this theory and study the spectrum of 
the linearization problem (\ref{linear1})--(\ref{linear4}) 
at the cut-off frequency $\lambda = \lambda_0(s)$, 
when $\Phi_u(r) = u_0(r)$ and $\Phi_w(r) = 0$. 
We will show that the vortex soliton is linearly stable 
in the neighborhood of the cut-off frequency 
$\lambda_0(s) \leq \lambda < \lambda_c(s)$, 
where $\lambda_c(s) \leq 1$ is the instability threshold. 
In the limit $\lambda \rightarrow \lambda_0(s)$, 
the linearization problem decomposes into two linear problems:
\begin{equation}
\label{partial1}
\pm i \sigma u_{\pm} = - u_{\pm}'' - \frac{1}{r} u_{\pm}' 
+ \left(1 + \frac{n^2}{r^2}\right) u_{\pm} 
- \frac{u_0^2}{1 + s u_0^2} u_{\pm} 
- \frac{u_0^2 (u_+ + u_-)}{(1 + s u_0^2)^2}, 
\end{equation}
and 
\begin{equation}
\label{partial2}
\pm i \sigma w_{\pm} = - w_{\pm}'' - 
\frac{1}{r} w_{\pm}' + (\lambda_0 + \frac{(n\mp 1)^2}{r^2}) 
w_{\pm} - \frac{u_0^2}{1 + s u_0^2} w_{\pm}.  
\end{equation}

The first linear problem (\ref{partial1}) 
is the stability problem of a scalar fundamental 
soliton $u = u_0(r)$ in a saturable medium.
The linear stability of such solitons has been well established
(see \cite{Yang02} for instance), thus unstable eigenvalues 
$\sigma$ do not exist in (\ref{partial1}). 
The continuous spectrum of the system (\ref{partial1}) 
is located at ${\rm Re}(\sigma) = 0$ and 
$|{\rm Im}(\sigma)| \geq 1$. The continuous 
spectrum is irrelevant for stability of solitary 
waves in the system of coupled NLS equations, 
if no embedded eigenvalues with negative energy 
quadratic form (\ref{quadform}) exist \cite{dmitry02}. 
The discrete spectrum of (\ref{partial1}) 
consists of isolated eigenvalues 
$\sigma$ such that ${\rm Re}(\sigma) = 0$ and $|{\rm Im}(\sigma)| < 1$, 
including the zero eigenvalue at $n = 0$ and $n= \pm 1$ 
with three eigenfunctions (\ref{zeroeigenfunction}) and 
three generalized eigenfunctions. 
Additional eigenvalues for internal modes exist in (\ref{partial1}) for 
${\rm Re}(\sigma) = 0$ and $0\ne |{\rm Im}(\sigma)|<1$.
These modes have been determined numerically 
in \cite{Yang02}. Using those numerical 
results, we have found that 
the energy quadratic form (\ref{quadform}) 
is positive for all internal modes of the system (\ref{partial1}). 
For instance, only one internal mode with $n=0$ exists 
and has positive value of $h$ 
for $s=0.5$ (which corresponds to $\omega=-0.5$ in \cite{Yang02}, 
see Fig. 3). We have also checked that no embedded eigenvalues 
with $|{\rm Im}(\sigma)| \geq 1$ 
exist in the problem (\ref{partial1}) for $s = 0.5$. 

The second linear problem (\ref{partial2}) is uncoupled 
for $w_+$ and $w_-$. Since the operator in the 
right-hand-side of Eq. (\ref{partial2}) 
is self-adjoint, the spectrum of $\sigma$ is purely 
imaginary, i.e. ${\rm Re}(\sigma) = 0$. The continuous spectrum of
(\ref{partial2}) is located at 
$|{\rm Im}(\sigma)| > \lambda_0$.  
Its discrete spectrum  
consists of isolated eigenvalues with 
${\rm Im}(\sigma) > - \lambda_0$ for $w_+$ 
and ${\rm Im}(\sigma) < \lambda_0$ for $w_-$ and 
can be embedded into continuous spectrum. 
The discrete spectrum of (\ref{partial2}) 
includes two zero eigenvalues 
at $n = 0$ with eigenfunctions $w_{\pm} = \phi(r)$, 
two zero eigenvalues at $n = \pm 2$ 
with eigenfunctions $w_{\pm} = \phi(r)$, and 
two non-zero eigenvalues $\sigma = \pm i (1 - \lambda_0)$ 
at $n = \pm 1$ with eigenfunctions $w_{\pm} = u_0(r)$. 
The zero eigenvalues at $n = 0$ are induced 
by the phase invariance of the $E_2$ equation, 
while the zero eigenvalues at $n = \pm 2$ are
induced by the symmetry of the uncoupled 
problem (\ref{partial2}). The eigenvalues
$\sigma = \pm i (1 - \lambda_0)$ with $n=\pm 1$ are induced 
by the arbitrary polarizations in the 
fundamental vector soliton (\ref{fundamental}) 
at $\lambda = \lambda_0(s)$. The latter eigenvalues 
result in negative values of 
the energy quadratic form (\ref{quadform}):
\begin{equation}
\label{negativeenergy}
h = - (1 - \lambda_0) \int_{0}^{\infty} u_0^2(r) r dr < 0.
\end{equation}
We have checked numerically that (\ref{partial2}) has no other
discrete eigenvalues for $s = 0.5$. 

Applying the stability theory of solitary waves \cite{dmitry02}, 
we count eigenvalues of the problems (\ref{partial1}) 
and (\ref{partial2}) that produce negative and zero values of 
the energy quadratic form $h$. At $\lambda = \lambda_0(s)$, 
only two eigenvalues $\sigma = \pm i (1 - \lambda_0)$ give 
the negative energy (\ref{negativeenergy}).  
Several zero eigenvalues give zero energy 
at $n = 0, \pm 1, \pm 2$. However, zero eigenvalues 
at $n = 0$ and $n = \pm 1$ 
are preserved at $\lambda > \lambda_0(s)$ due to 
translation, rotation, and complex-phase symmetries of the system 
(\ref{E1})--(\ref{E2}). Only two zero eigenvalues of the problem 
(\ref{partial2}) at $n = \pm 2$ are not preserved by the symmetry 
and they can move out of zero for $\lambda>\lambda_0(s)$. 
We shall now consider the shift of these negative-energy and zero-energy 
eigenvalues for $\lambda$ near the cut-off frequency $\lambda_0(s)$. 

We show first that the negative-energy eigenvalues $\sigma = \pm i (1 - \lambda_0)$ 
never bifurcate off the imaginary axis 
for $\lambda > \lambda_0(s)$ regardless whether they are embedded or not.  
Indeed, at any $\lambda$ value, the 
linearization problem (\ref{linear1})--(\ref{linear4}) 
has the exact discrete eigenmode  
\begin{equation}
\label{exactmode1}
u_+ = 0, \quad u_- = - \Phi_w(r), \quad
w_+ = \Phi_u(r), \quad w_- = 0
\end{equation}
for $n = 1$ and $\sigma = i ( 1 - \lambda)$, 
and 
\begin{equation}
\label{exactmode2}
u_+ = - \Phi_w(r), \quad u_- = 0, \quad
w_+ = 0, \quad w_- = \Phi_u(r)
\end{equation}
for $n=-1$ and $\sigma = -i ( 1 - \lambda)$. 
This result contrasts what happens in the system of 
coupled NLS equations with Kerr nonlinearities, where 
the negative-energy discrete eigenvalues, 
which are embedded in the continuous spectrum, bifurcate to the complex plane
and lead to the instability \cite{dmitryYang,dmitry02}. 

We note that the exact eigenmodes (\ref{exactmode1}) and (\ref{exactmode2}) 
generate an approximate solution of the system 
(\ref{E1})--(\ref{E2}):
\begin{equation} \label{E1E2solution}
E_1 = \Phi_u(r) e^{iz} - \gamma \Phi_w(r) e^{i \theta + i \lambda z} 
+ {\rm O}(\gamma^2), \quad 
E_2 = \Phi_w(r) e^{i \theta + i \lambda z} + \gamma 
\Phi_u(r) e^{iz} + {\rm O}(\gamma^2),
\end{equation}
where $\gamma$ is an arbitrary small parameter. 
Solution (\ref{E1E2solution}) is nothing but the original vortex vector soliton
under a small rotation in the $(E_1, E_2)$ plane. This rotation leaves the
original system (\ref{E1}) and (\ref{E2}) invariant and is one of the symmetries of
the present problem. Thus, although solution (\ref{E1E2solution})
appears like internal oscillations of vortex solitons, this oscillation
does not create any energy radiation and is fundamentally different from
internal oscillations discussed in \cite{Yang02,rosanov,dmitryYang}. 

\begin{figure}
\setlength{\epsfxsize}{6cm}\epsfbox{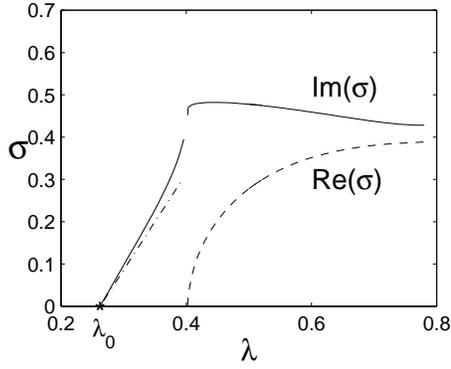}
\caption{Eigenvalues $\sigma$ of vortex solitons versus $\lambda$ at $s=0.5$ and $n=2$.
The cut-off frequency $\lambda_0$ is marked by '*'. Solid line: Im($\sigma$); 
dashed line: Re($\sigma$); dash-dotted line: analytical formula 
(\ref{perturbationUW}) and (\ref{sigma2}).  }
\label{eigenvalue}
\end{figure}

We show next that the zero eigenvalues at $n=\pm 2$ 
move to the imaginary axis (as conjugate pairs) as $\lambda > \lambda_0(s)$ 
and do not create any instability. 
We will use the perturbation 
series expansions and will present calculations only for the case 
$n = 2$ (the case $n = -2$ is similar).
When $\lambda$ is close to $\lambda_0(s)$, 
we construct an approximate solution to 
the linearization problem (\ref{linear1})--(\ref{linear4}) 
at $n = 2$ in the form:
\begin{equation}   
\label{perturbationUW}
u_{\pm} = \epsilon u_{\pm}^{(1)}(r) + {\rm O}(\epsilon^3), \quad 
w_+ = \phi(r) + \epsilon^2 w_+^{(2)}(r) + {\rm O}(\epsilon^4), \quad 
w_- = \epsilon^2 w_-^{(2)}(r) + {\rm O}(\epsilon^4), \quad
\sigma = \epsilon^2 \sigma_2 + {\rm O}(\epsilon^4), 
\end{equation}
where $\epsilon$ is the same small parameter as in 
expansions (\ref{uexpand})--(\ref{lambdaexpand}). 
Substituting (\ref{perturbationUW}) into the system 
(\ref{linear1})--(\ref{linear4}), we find an exact 
solution at order $\epsilon$: 
$u_+^{(1)} = u_-^{(1)} = u_{22}(r)$, 
where $u_{22}(r)$ solves the equation (\ref{u22}). 
At order $\epsilon^2$, we need to solve 
the non-homogeneous equation for $w_+^{(2)}(r)$:
\begin{eqnarray} \label{wCorrection}
w_+^{(2)\prime\prime} + \frac{1}{r} w_+^{(2)\prime\prime} 
- \left( \lambda_0 + \frac{1}{r^2} \right) w_+^{(2)} 
+ \frac{u_0^2}{1 + s u_0^2} w_+^{(2)} = \left( \lambda_2 - i \sigma_2 \right) \phi 
- \frac{2 \phi (u_0 u_{20} + u_0 u_{22} + \phi^2)}{(1 + s u_0^2)^2}.
\end{eqnarray}
The solvability condition for this equation can be simplified 
by virtue of Eq. (\ref{vortexlambda2}), and we find that the
eigenvalue coefficient $\sigma_2$ is given as
\begin{equation}
\label{vortexsigma2}
\sigma_2 = 2 i \frac{\int_0^{\infty} r \phi^2 h_2 dr}{
\int_0^{\infty} r \phi^2 dr}.
\end{equation}
Utilizing Eq. (\ref{integral}), 
we see that
\begin{equation} \label{sigma2}
\sigma_2=2i \left[ 2\lambda_{2d}(s)-\lambda_{2v}(s) \right], 
\end{equation}
whose imaginary part is positive from Fig. \ref{lambda02}a.
Thus, the energy quadratic form of the bifurcated eigenmode (\ref{perturbationUW}) and (\ref{sigma2}) 
(up to the order $\epsilon^2$) is negative: 
\begin{equation}
\label{negativeenergy2}
h = - \epsilon^2 {\rm Im}(\sigma_2) \int_0^{\infty} r \phi^2 dr < 0. 
\end{equation}
The analytical eigenvalue formula (\ref{perturbationUW}) and (\ref{sigma2})
at $n = 2$ is plotted in Fig. \ref{eigenvalue} versus $\lambda$ for $s=0.5$ (dash-dotted line). 
Numerically, we have determined these eigenvalues for $s=0.5$ and 
various values of $\lambda$, and the results are plotted in Fig. \ref{eigenvalue} (solid line) as well. 
When $\lambda$ is close to $\lambda_0$, 
the analytical formula agrees well with the numerical values. 

We have shown above that the two zero eigenvalues of 
the system (\ref{partial2}) at $n=\pm 2$ 
move to the imaginary axis when $\lambda>\lambda_0(s)$, while the two 
non-zero negative-energy eigenvalues at $n = \pm 1$ 
remain on the imaginary axis. Thus, 
we conclude that vortex solitons 
are linearly stable near the cut-off frequency $\lambda=\lambda_0(s)$, i.e. 
vortex solitons with small vortex components are linearly stable. 
This result confirms the conclusions of \cite{ziad00}
and does not support conclusions of \cite{garcia} where 
{\em all} vortex vector solitons were claimed to be linearly
unstable.  

\begin{figure}
\begin{tabular}{cc}
\setlength{\epsfxsize}{3cm}\epsfbox{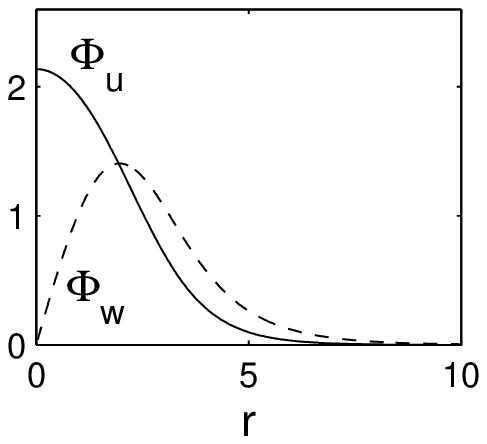}
\setlength{\epsfxsize}{5cm}\epsfbox{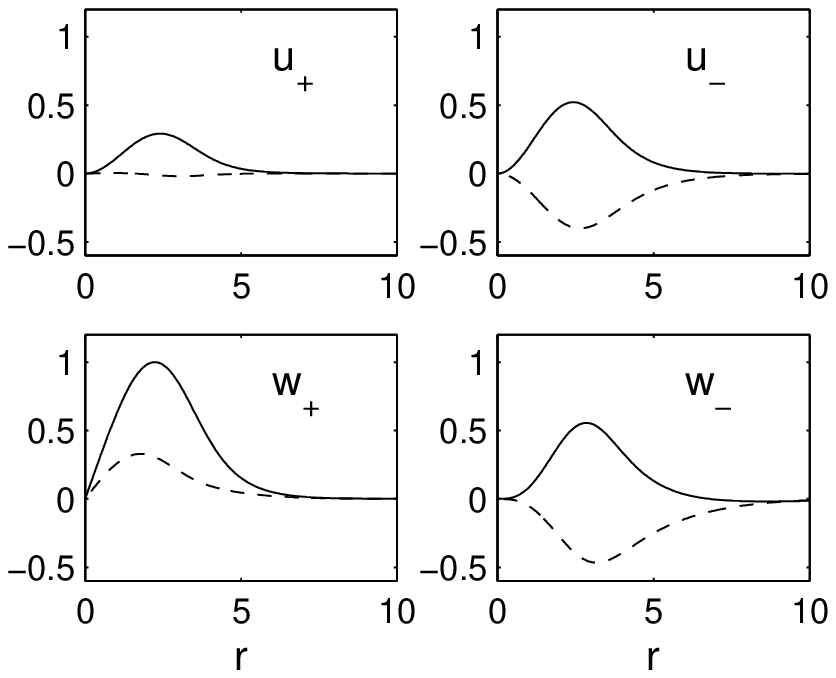}
\end{tabular}
\caption{The vortex soliton (left) and its unstable eigenmode (right) at $s=0.5$, $\lambda=0.5$ and $n=2$. 
In the right figure, solid lines are the real parts of the eigenfunctions, and dashed lines are the
imaginary parts.}
\label{unstablemodes}
\end{figure}

Unstable eigenvalues of vortex solitons may appear 
far away from the cut-off frequency $\lambda_0(s)$. 
Indeed, the two imaginary eigenvalues $\sigma$ for 
$n = \pm 2$ that bifurcate from zero eigenvalues 
at $\lambda > \lambda_0(s)$ have negative energy 
(\ref{negativeenergy2}). 
When these eigenvalues collide with eigenvalues of positive energy 
or with continuous spectrum, the oscillatory instability may arise \cite{skryabin,dmitry02}. 
We confirm this scenario and compute unstable eigenvalues $\sigma$ 
of the linear system (\ref{linear1})--(\ref{linear4}) with the numerical shooting method. 
The unstable eigenvalues are found exactly at $n = \pm 2$ and 
are shown in Fig. \ref{eigenvalue} for $s=0.5$.  
The unstable eigenvalues appear when  
$\lambda > \lambda_c \approx 0.402$, where $\lambda_c$ 
denotes the frequency for onset of instability. 
These results agree with Fig. 3 of Ref. \cite{garcia}, where the 
unstable eigenvalues were found from time-integration of 
the linearized equations under random-noise initial perturbations. 
Fig. \ref{unstablemodes} displays our numerical solutions 
for the vortex soliton and 
the corresponding unstable eigenfunction  
at $s=0.5$,  $\lambda=0.5$ and $n=2$. 
Thus, for the case $s = 0.5$, 
unstable eigenvalues exist at $\lambda > \lambda_c \approx 0.402$, 
while vortex vector solitons exist at 
$\lambda > \lambda_0 \approx 0.2622$, see Fig. \ref{eigenvalue}. 
In the interval $\lambda_0 < \lambda < \lambda_c$, i.e., 
$0.2622 < \lambda < 0.402$ for $s = 0.5$, 
unstable eigenvalues do not exist and the vortex 
solitons are linearly {\em stable}. 

We conclude this analysis with two remarks. 
First, it follows from Fig. \ref{eigenvalue} for 
$s = 0.5$ that 
the eigenvalues $\sigma$ at $n=\pm 2$
merge into the continuous spectrum at $\lambda \approx 0.396$, while 
unstable eigenvalues appear at $\lambda=\lambda_c \approx 0.402$. 
Our numerical results are inconclusive as to what 
happens in the narrow interval $0.396 < \lambda < 0.402$. 
This problem is left open for future studies. 
And second, when $\lambda$ is further away 
from the cut-off frequency $\lambda_0(s)$,
the vector vortex solution bifurcates into scalar vortex solutions with 
$u = 0$ and $w = \Phi_w(r) e^{i \theta}$, 
see \cite{desyatnikov}. The scalar vortex soliton 
has additional unstable eigenmodes at $|n|\ne 2$ which have
smaller growth rates (see \cite{firthskryabin}). 
We do not study this bifurcation 
where the family of vector vortex solitons terminates, 
nor the number of unstable eigenvalues 
of vector vortex solitons near this bifurcation.

\subsection{Dipole soliton}

To study the linear stability of the dipole solitons 
(\ref{dipole}), we linearize the system 
(\ref{E1})--(\ref{E2}) with the perturbation:
\begin{equation}
E_1 = e^{iz} \left[ U(x,y) + 
\left( u_r(x,y) + u_i(x,y) \right) e^{\sigma z} + 
\left( \bar{u}_r(x,y) - \bar{u}_i(x,y) \right) 
e^{\bar{\sigma} z} \right], 
\end{equation}
\begin{equation}
E_2=e^{i\lambda z} \left[ W(x,y)+ 
\left( w_r(x,y) + w_i(x,y) \right) e^{\sigma z} + 
\left( \bar{w}_r(x,y) - \bar{w}_i(x,y) \right) 
e^{\bar{\sigma} z} \right]. 
\end{equation}
Here $u_r, u_i, w_r$ and $w_i$ are complex functions and are very small. 
The linearization problem is then written in the form:
\begin{eqnarray}
\label{linear11}
i \sigma u_i & = & - \Delta u_r + u_r 
- \left( V + 2 V_{uu} \right) u_r - 2 V_{uw} w_r, \\
\label{linear12}
i \sigma u_r & = & - \Delta u_i + u_i - V u_i, \\
\label{linear13}
i \sigma w_i & = & - \Delta w_r + \lambda w_r 
- \left( V + 2 V_{ww} \right) w_r - 2 V_{uw} u_r, \\
\label{linear14}
i \sigma w_r & = & - \Delta w_i + \lambda w_i - V w_i, 
\end{eqnarray}
where 
$$
V = \frac{U^2 + W^2}{1 + s(U^2 + W^2)}, \quad
V_{uu} = \frac{U^2}{(1 + s(U^2 + W^2))^2}, \quad 
V_{uw} = \frac{U W}{(1 + s(U^2 + W^2))^2}, \quad
V_{ww} = \frac{W^2}{(1 + s(U^2 + W^2))^2}.
$$
The linearized problem can be formulated in the same 
Hamiltonian form (\ref{Lphi}) with the energy 
quadratic form \cite{dmitry02}:
\begin{equation}
\label{quadform1}
h = i \sigma \int_{-\infty}^{\infty}\!\!\int_{-\infty}^{\infty} 
\left( \bar{u}_r u_i + \bar{u}_i u_r + \bar{w}_r w_i + 
\bar{w}_i w_r \right) dx dy. 
\end{equation} 
At the cut-off frequency $\lambda = \lambda_0(s)$, 
the same analysis as for the vortex solitons shows 
existence of a pair of eigenvalues $\sigma = \pm i (1 - \lambda_0)$ 
with negative values of $h$ and a number of zero eigenvalues 
with zero values of $h$. We show again that the eigenvalues 
$\sigma = \pm i (1 - \lambda_0)$ with negative energy 
never bifurcate into complex domain 
for $\lambda > \lambda_0(s)$. Indeed, for any $\lambda$ value,  
the linearization problem (\ref{linear11})--(\ref{linear14}) 
has the exact solution 
\begin{equation}
\label{embeddedeigenvalue1}
u_r = -W(x,y), \quad u_i = W(x,y), \quad
w_r = U(x,y), \quad w_i = U(x,y)
\end{equation}
at $\sigma = i ( 1 - \lambda)$, and 
\begin{equation}
\label{embeddedeigenvalue11}
u_r = W(x,y), \quad u_i = W(x,y), \quad
w_r = -U(x,y), \quad w_i = U(x,y)
\end{equation}
at $\sigma = -i ( 1 - \lambda)$. 

We study the zero eigenvalues of the system (\ref{linear11})--(\ref{linear14}) 
with perturbation series expansions for $\lambda > \lambda_0(s)$:
\begin{eqnarray}
\nonumber
\sigma = \epsilon^2 \sigma_2 + {\rm O}(\epsilon^4), \quad
u_r = \epsilon u_r^{(1)}(r,\theta) + {\rm O}(\epsilon^3), \quad 
u_i = {\rm O}(\epsilon^3), \\
\label{perturbationUW1} 
w_r = w_r^{(0)}(r,\theta) + \epsilon^2 w_r^{(2)}(r,\theta) 
+ {\rm O}(\epsilon^4), \quad
w_i = w_i^{(0)}(r,\theta) + \epsilon^2 w_i^{(2)}(r,\theta) 
+ {\rm O}(\epsilon^4).
\end{eqnarray}
Here $\epsilon$ is the same small parameter as in 
expansions (\ref{uexpand})--(\ref{lambdaexpand}), 
while the functions $w_{r,i}^{(0)}(r,\theta)$ 
are linear combinations of the eigenfunctions of 
the null space of the problem (\ref{linear11})--(\ref{linear14}) 
at $\lambda = \lambda_0(s)$:
\begin{equation}
\label{linearcombination}
w_r^{(0)} = c_1 \phi(r) \cos \theta + c_2 \phi(r) \sin \theta, \quad 
w_i^{(0)} = d_1 \phi(r) \cos \theta + d_2 \phi(r) \sin \theta,
\end{equation}
where $c_1$, $c_2$, $d_1$, and $d_2$ are constants. 
Substituting (\ref{perturbationUW1}) into the system 
(\ref{linear11})--(\ref{linear14}), we find an exact 
solution at order $\epsilon$: 
\begin{equation}
\label{firstU}
u_r^{(1)} = c_1 u_{20}(r) + \left( c_1 \cos 2 \theta 
+ c_2 \sin 2 \theta \right) u_{22}(r).
\end{equation} 
where $u_{20}(r)$ and $u_{22}(r)$ solve the problems 
(\ref{u20}) and (\ref{u22}). 
At order $\epsilon^2$, four solvability conditions 
are needed for solving the non-homogeneous equations 
for $w_r^{(2)}(r,\theta)$ and $w_i^{(2)}(r,\theta)$. 
Using Eq. (\ref{lambda2}), we transform the four solvability 
conditions to the form:
\begin{eqnarray}
\label{dipolesigma2}
\sigma_2 c_1 = 0, \quad \sigma_2 d_2 = 0, \quad
i \sigma_2 c_2 \int_0^{\infty} r \phi^2 dr = 
d_2 \int_0^{\infty} r \phi^2 h_2 dr, \quad
i \sigma_2 d_1 \int_0^{\infty} r \phi^2 dr = - c_1 
\int_0^{\infty} r \phi^2 \left( 2 h_1 + h_2 \right) dr.
\end{eqnarray}
If $\sigma_2 = 0$, then $c_1 = d_2 = 0$, while 
$c_2$, $d_1$ are arbitrary constants. Thus, the zero 
eigenvalue persists in the system (\ref{linear11})--(\ref{linear14}) 
for $\lambda > \lambda_0(s)$ with two eigenfunctions 
$w_r = \phi(r) \sin\theta$ and $w_i = \phi(r) \cos \theta$. 
The two eigenfunctions are related to the symmetries of 
the system (\ref{E1})--(\ref{E2}) with respect to 
rotation in $\theta$ and shift of the complex phase. 
If $\sigma_2 \neq 0$ however, the system (\ref{dipolesigma2}) 
has only the trivial solution: $c_1 = c_2 = d_1 = d_2 = 0$. 
Therefore, the other two zero eigenvalues do not 
bifurcate to the imaginary axis but simply disappear 
for $\lambda > \lambda_0(s)$. 

We have analytically proved above that the dipole solitons 
are linearly stable in the neighborhood of the 
cut-off frequency $\lambda_0(s)$. Moreover, 
contrary to vortex solitons, there are 
only two eigenvalues of negative energy 
for $\lambda > \lambda_0(s)$, and they remain on the imaginary axis for
all $\lambda$ values [see (\ref{embeddedeigenvalue1}) and (\ref{embeddedeigenvalue11}]. 
Thus, we conjecture that 
the dipole solitons are linearly stable 
in the whole domain of their existence. 
This conjecture is in agreement with the numerical work of \cite{garcia}. 
We again confirm this result by numerical 
simulations of the system (\ref{E1})--(\ref{E2}) 
linearized around the dipole soliton (\ref{dipole}). 
For $s=0.5$, we have simulated the linearized system 
for several values of $\lambda$ between 
$\lambda = 0.3$ and  $\lambda = 0.85$. 
We did not find any instability in the linearized 
system. Since $\lambda=0.3$ is close to the cut-off frequency 
$\lambda_0=0.2622$ and $\lambda=0.85$ is close to the 
end frequency $\lambda = 1$, we conclude that 
dipole solitons are indeed linearly {\em stable} 
in the whole existence interval. 

\section{Nonlinear evolution of perturbed vortex solitons}
\label{Secevolution}

Here we study the nonlinear evolution 
of perturbed vortex solitons. The unstable 
vortex soliton under small random-noise perturbations 
was found in \cite{garcia} to break up into a rotating
dipole vector soliton. We will show below that such a breakup scenario
holds only when the vortex component of the vortex soliton is below certain
threshold. Above that threshold, unstable vortex solitons
break up into two rotating fundamental vector solitons instead. We will also show that 
the vortex solitons with small vortex components are 
not only linearly but also nonlinearly stable.   

\begin{figure}
\setlength{\epsfxsize}{4.5cm}\epsfbox{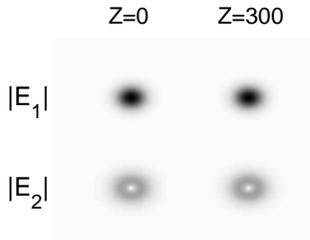}
\caption{Stable evolution of vortex vector solitons with $\lambda < \lambda_c$ 
under the perturbation (\ref{icvortex}) with $s=0.5$, $\lambda=0.38$, and 
$\alpha=0.05$. }
\label{stablevortex}
\end{figure}

We consider first the nonlinear evolution of linearly stable vortex solitons. 
For this purpose, we have simulated the system (\ref{E1})--(\ref{E2})
starting with a linearly-stable vortex soliton under various types of small initial perturbations
such as random-noise and deterministic ones. We have found that 
the vortex solitons are also nonlinearly stable for all small perturbations. 
To demonstrate, we select $s=0.5$ and $\lambda=0.38$, where the vortex soliton
has been shown to be linearly stable (see Fig. \ref{eigenvalue}). 
As initial perturbations, we choose 
\begin{equation}\label{icvortex}
E_1(r, \theta, 0)=(1+\alpha)\Phi_u(r,\lambda), \quad 
E_2(r, \theta, 0)=(1+\alpha)\Phi_w(r,\lambda) e^{i\theta}, 
\end{equation}
where $\alpha$ is a small perturbation parameter, 
that measures amplification of the vortex soliton
by a factor $1+\alpha$. The simulation result with $\alpha=0.05$ is shown in 
Fig. \ref{stablevortex}. 
This figure shows that the perturbed vortex soliton 
persists the nonlinear evolution 
and exhibits little change of shape even after 300 
diffraction lengths. This clearly 
confirms the linear and nonlinear stability of the vortex soliton with $s=0.5$ and $\lambda=0.38$.  
Other perturbations to this soliton give similar evolution results. 

We study next the nonlinear evolution of linearly unstable vortex solitons.
We have shown in Sec. \ref{section3a} that these solitons possess two 
unstable eigenmodes $(\sigma, n, u_+, u_-, w_+, w_-)$ and 
$(\bar{\sigma}, -n, \bar{u}_-, \bar{u}_+, \bar{w}_-, \bar{w}_+)$ with $n=2$. 
The $\sigma$ versus $\lambda$ graph is shown in Fig. \ref{eigenvalue} for $s = 0.5$, 
while unstable eigenfunctions $(u_+, u_-, w_+, w_-)$ for $s = 0.5$ and $\lambda = 0.5$
are displayed in Fig. \ref{unstablemodes}.
However, we recognize that these two unstable eigenmodes are equivalent in view of 
Eqs. (\ref{E1perturb}) and (\ref{E2perturb}). 
Thus, any small initial perturbation to the vortex soliton is projected onto 
this unstable eigenmode which grows exponentially, while the rest of the initial perturbation
disperses away.
For convenience, we choose the initial perturbation 
to be exactly this unstable eigenmode, i.e., 
\begin{equation} \label{ic2a}
E_1(r,\theta, 0) = \Phi_u(r) + \alpha(
u_+(r) e^{- 2 i\theta} + 
\bar{u}_-(r) e^{2 i \theta}), 
\end{equation}
\begin{equation}\label{ic2b}
E_2(r,\theta,0)=e^{i\theta} \left[ \Phi_w(r)+ 
\alpha(w_+(r) e^{-2 i\theta} + 
\bar{w}_-(r) e^{2 i \theta }) \right], 
\end{equation}
where $\alpha$ is a small perturbation parameter. The advantage of this special 
perturbation is that it shortens the distance
for the breakup of the vortex soliton and reduces the radiation 
noise in the nonlinear evolution of the perturbed solution. 

\begin{figure}
\setlength{\epsfxsize}{9cm}\epsfbox{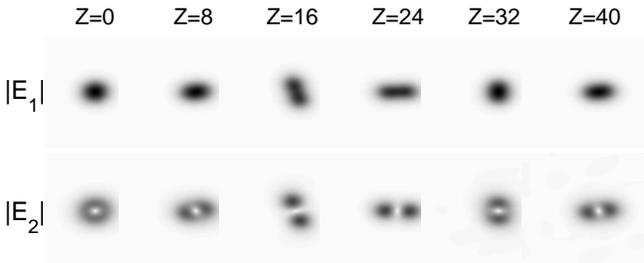}
\caption{Break-up of an unstable vortex soliton into a rotating dipole soliton 
under the perturbation (\ref{ic2a})--(\ref{ic2b}) with $s=0.5$, $\lambda=0.45$, 
and $\alpha=0.05$.}
\label{unstablevortex1}
\end{figure}

\begin{figure}
\setlength{\epsfxsize}{8cm}\epsfbox{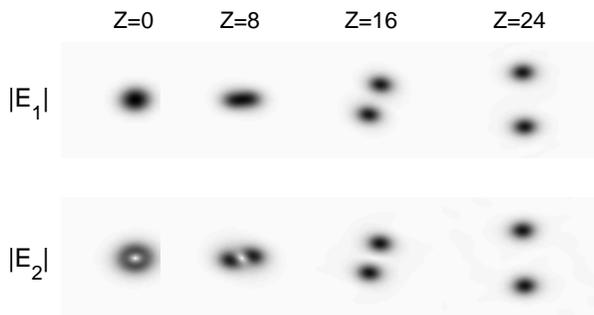}
\caption{Break-up of an unstable vortex soliton into two fundamental solitons 
under the perturbation (\ref{ic2a})--(\ref{ic2b}) with $s=0.5$, $\lambda=0.5$, 
and $\alpha=0.05$.}   
\label{unstablevortex2}
\end{figure}

We have discovered two breakup scenarios of the unstable vortex soliton 
with the initial perturbation (\ref{ic2a})--(\ref{ic2b}). 
We confirm that unstable vortex solitons with relatively {\em small} vortex components indeed
break up into a rotating dipole soliton,  in agreement with \cite{garcia}. 
However, when the vortex component increases above a certain threshold, an unstable vortex soliton
breaks up into two rotating {\em fundamental} vector solitons rather than one dipole soliton. 
For example, when $s=0.5$ and $\alpha = 0.05$, the vortex soliton breaks up into a dipole soliton when
$0.402 < \lambda \: _{\sim}^{<} \: 0.45$, and into two fundamental vector solitons when
$\lambda > 0.45$. Indeed, when $\lambda=0.45$ (where the vortex component is relatively small),  
the time evolution of the perturbed vortex soliton 
is plotted in Fig. \ref{unstablevortex1}. It is seen that 
this soliton breaks up into a rotating dipole soliton. 
But when $\lambda=0.5$ (where the vortex component is bigger), 
the time evolution is shown in Fig. \ref{unstablevortex2}. Here
two rotating fundamental vector solitons are formed after the breakup 
of the unstable vortex soliton. 
We have also found that these breakup scenarios are insensitive to
the type of initial perturbation imposed, because we have simulated
the evolutions with different values of $\alpha$ in 
(\ref{ic2a}) and (\ref{ic2b}) as well as with other forms of initial perturbations
such as random noise, but the breakup scenarios do not change.
To check the numerical accuracy of our simulations, we have used
more grid points and wider $(x,y)$ intervals and obtained identical results. 
Furthermore, our results conserve energies of the $E_1$ and $E_2$ components very well. 

Intuitively, it is not difficult to understand the above two breakup scenarios of
unstable vortex solitons. When the vortex component of the vortex soliton is small, 
the instability (with $n=2$) breaks up the vortex ($E_2$) component into two weak humps, 
while it does not significantly affect the single-hump shape of the 
fundamental ($E_1$) component since $E_1$'s initial amplitude
is much higher. During the subsequent evolution, 
the two humps of the $E_2$ component are too weak to break the $E_1$ component
into two pieces, thus the solution relaxes into a dipole soliton instead of 
two fundamental solitons. However, when the vortex component of the vortex soliton
is sufficiently large, the fundamental component becomes small 
(see Fig. \ref{vortexfig} and \cite{garcia}). In this case, instability breaks up
both the vortex and fundamental components into two pieces, and two fundamental solitons 
are formed then. 

\section{Summary and discussion}

To summarize, we have studied both analytically 
and numerically the existence, uniqueness, and 
stability of vortex and dipole vector solitons 
in saturable optical materials in (2+1) dimensions. 
We have shown that 
the analytical expressions for vortex and 
dipole vector solitons can be constructed 
with perturbation series expansions near the 
cut-off frequency $\lambda = \lambda_0(s)$. 
We have also shown that only two vector solitons 
bifurcate from the same cut-off frequency, which 
are vortex and dipole solitons. Furthermore, 
we have proved that both vortex and dipole solitons 
are linearly {\em stable} when the vortex and dipole
components are {\em small}. As the vortex and 
dipole components increase, the family of 
vortex vector solitons becomes linearly unstable, 
while that of dipole vector solitons remains linearly stable 
in the entire existence domain. We have also shown that 
unstable vortex solitons break up into a rotating dipole soliton
only when the vortex component is relatively small. 
When the vortex component crosses a certain threshold, 
the vortex soliton breaks up into two rotating 
fundamental vector solitons instead. 
We expect that our results are significant not only for 
studies of spatial vector solitons in a saturable nonlinear medium
but also for studies of Bose-Einstein condensation.  

In this paper, we have studied only the simplest vortex and dipole vector 
solitons which bifurcate from the fundamental $u$ and small $w$ components.
One natural question to ask is the existence and stability of other vortex
and multi-pole vector solitons. The perturbation 
series expansion method developed in this paper is powerful  
for a systematic study of general vortex and multi-pole vector solitons 
near their bifurcation points. But this problem lies outside the scope of
the present article. We note, however, that vortex solitons (\ref{vortex})
with $|n|>0$ and $|m|>0$ exist, and they are expected to be linearly always
unstable because each component has non-zero charge and is linearly unstable
by itself \cite{firthskryabin}.
This expectation is consistent with our preliminary numerical simulations
on vortex solitons with charges such as $n=1$ and $m=-1$.

Recently, three-component vortex and dipole vector solitons in a saturable medium
have been investigated \cite{kivshar02}. The authors found that
those solitons are linearly unstable provided their total topological charge is 
non-zero. In view of our results in this paper, that conclusion needs
modification. We plan to study that system carefully in the near future.

\section*{\hspace{0.1cm} Acknowledgments}

The authors appreciate helpful discussions with Yu. Kivshar, 
Z. Musslimani, B. Sandstede and D. Skryabin. 
The work of D.P. was supported in part by 
the NSERC grant No. 5-36694 and CFI grant No. 5-26773. 
The work of J.Y. was supported in part by 
the Air Force Office of Scientific Research under 
contract F49620-99-1-0174, and by the National Science 
Foundation under grant DMS-9971712. 

\end{document}